\documentclass[10pt, onecolumn]{article}
\usepackage[breaklinks=true]{hyperref}
\usepackage{amsmath}
\usepackage{authblk}
\usepackage{float}
\usepackage[margin=1in]{geometry}
\usepackage{graphicx}
\usepackage{tikz}
\usepackage[utf8]{inputenc}
\usepackage[numbers]{natbib}
\usepackage{tabularx}
\usepackage{url}
\newcolumntype{n}{X}
\newcolumntype{s}{>{\hsize=.35\hsize}X}

\restylefloat{table}

\title{Google COVID-19 Search Trends Symptoms Dataset: Anonymization Process Description (version 1.0)}

\author{Shailesh Bavadekar}
\author{Andrew Dai}
\author{John Davis}
\author{Damien Desfontaines}
\author{Ilya Eckstein}
\author{Katie Everett}
\author{Alex Fabrikant}
\author{Gerardo Flores}
\author{Evgeniy Gabrilovich}
\author{Krishna Gadepalli}
\author{Shane Glass}
\author{Rayman Huang}
\author{Chaitanya Kamath}
\author{Dennis Kraft}
\author{Akim Kumok}
\author{Hinali Marfatia}
\author{Yael Mayer}
\author{Benjamin Miller}
\author{Adam Pearce}
\author{Irippuge Milinda Perera}
\author{Venky Ramachandran}
\author{Karthik Raman}
\author{Thomas Roessler}
\author{Izhak Shafran}
\author{Tomer Shekel}
\author{Charlotte Stanton}
\author{Jacob Stimes}
\author{Mimi Sun}
\author{Gregory Wellenius}
\author{Masrour Zoghi}
\affil{covid-19-search-trends-feedback@google.com}

\begin{document}
\maketitle
\begin{abstract}
This report describes the aggregation and anonymization process applied to the initial version of COVID-19 Search Trends symptoms dataset~\cite{symptomdataset}, a publicly available dataset that shows aggregated, anonymized trends in Google searches for symptoms (and some related topics). The anonymization process is designed to protect the daily symptom search activity of every user with $\varepsilon$-differential privacy for $\varepsilon$ = 1.68.
\end{abstract}

\vfill

\section{Introduction}

The COVID-19 Search Trends symptoms dataset is a publicly available dataset that shows aggregated, anonymized trends in Google searches for symptoms (and some related topics). The dataset provides a daily or weekly time series for each region showing the relative volume of searches for each symptom. Its goal is to help researchers, public health experts and data analysts better understand the impact of COVID-19, while protecting our users’ privacy. For this purpose, the dataset covers approximately 400 symptoms, such as cough, fever and difficulty breathing. To ensure strict privacy standards, all published data is aggregated and anonymized, protecting each user’s symptom search activity on a given day using a differentially private mechanism. No personal data or individual searches are included in the dataset.

The data of the symptoms dataset reflect the volume of Google searches that may be associated with a particular medical symptom. For this purpose, we count the daily number of searches relating to that symptom within a given geographic region and normalize this number based on the total search activity in that region. The resulting dataset is either a daily or a weekly series showing the relative frequency of searches for a particular symptom in a particular region. The dataset covers the recent period and we’ll gradually expand its range as part of regular updates.
 
Similar to the Google COVID-19 Community Mobility Reports~\cite{mobilitydataset, mobilitytechreport} and as explained in greater technical detail below, the anonymization process of the symptoms dataset is based on differential privacy~\cite{dp}, which is a well established concept for producing data that satisfy formal privacy guarantees. For this purpose, we intentionally perturb our data by adding random noise and drop data that is deemed to be unreliable. The symptoms dataset is designed to maintain the privacy of our users while releasing aggregated and anonymized data that is as accurate and useful as possible. 

The remainder of this report is structured as follows: First, we introduce some basic concepts and terminology of the symptoms dataset. We then explain how differential privacy is used to produce anonymized aggregates. Finally, we elaborate on how the published data is built from anonymized aggregates.

\section{Definitions}
The following definitions explain some common terms and concepts that we use throughout this report.

\paragraph{User}
A user who did a web search on Google.

\paragraph{Symptom Search}
A symptom search is a Google search query issued by a search user that relates to a particular medical symptom or health condition. The symptoms we are considering are compiled in a predefined list including medical issues such as cough, fever and difficulty breathing.

\paragraph{Geographic Granularity}
The COVID-19 Search Trends symptoms dataset is aggregated per geographic region. Similarly to the Google COVID-19 Community Mobility Reports~\cite{mobilitydataset, mobilitytechreport}, we distinguish between three levels of geographic regions, which we call granularity levels:

\begin{itemize}
    \item Granularity level 0 corresponds to data aggregated by country.
    \item Granularity level 1 corresponds to data aggregated by top-level geopolitical subdivisions (e.g., US states or equivalent geographical regions in other countries).
    \item Granularity level 2 corresponds to data aggregated by higher-resolution granularity (e.g., US counties or equivalent geographical regions in other countries).
\end{itemize}

Granularity levels 1 and 2 are defined differently in different countries to account for the differences between countries’ public health systems. Note that in general, the area of a geographic region gets smaller as the granularity number increases. All regions of the symptoms dataset have an area of at least 3km$^2$.

\paragraph{Temporal Granularity}
For each geographic region and symptom, the data is released either as daily or weekly aggregates. The particular temporal granularity depends on the quality of the data. In general, we try to provide daily aggregates whenever possible. However, if the accuracy of our data is affected too much by our privacy protections, we may choose to release weekly aggregates instead. Weekly aggregates help to improve accuracy while maintaining the same level of privacy. Since weekly aggregates are based on more data points, the noise added for differential privacy introduces less relative error.

The decision between daily and weekly aggregates is made per symptom and region based on the data available at the time of the initial release of the symptoms dataset, and is made in a differentially private manner (more details can be found in the final paragraph of this report). We keep the temporal granularity fixed for a given region and symptom throughout the full time range of the dataset release.

\paragraph{Differential Privacy}
Let $A$ be a randomized algorithm for computing some metric over a given dataset. In the context of this report, we consider a pair of datasets $D_1$ and $D_2$ neighboring if $D_2$ can be obtained from $D_1$ by adding or removing a single user’s search activity on a given day. We then call $A$ $\varepsilon$-differentially private if for any pair of neighboring datasets $D_1$ and $D_2$ and for all sets $S$ of the possible outputs of $A$\footnote{We generate a fixed set of metrics (all $<$day, symptom, region$>$ combinations) and add Laplace noise to all of them independent of whether they contain user data or not. Thus, the released metrics satisfy ($\varepsilon$, $\delta$)-differentially privacy, with $\delta=0$.}:
\[\textrm{Pr}[A(D_1) \in S] \leq \exp(\varepsilon) \cdot \textrm{Pr}[A(D_2) \in S].\]

\section{Anonymizing User Data}
\begin{figure}
    \centering
    \begin{tikzpicture}[x=0.65pt,y=0.65pt,yscale=-1,xscale=1]
        %Dashed Line
        \draw [color={rgb, 255: red, 0; green, 0; blue, 0}, draw opacity=1] [dash pattern={on 0.45pt off 3.0pt}] (386.01,10.19) -- (386.01,268.56);
        \node[align=center, scale=0.6] at (426.01,263.19) {privacy boundary};
        %Polygon 
        \draw [fill=white] (514.51,26.31) -- (514.51,154.31) -- (375.51,154.31) -- (375.51,144.31) -- (406.51,116.31) -- (406.51,26.31) -- cycle;
        %Rectangles
        \draw [fill=white, fill opacity=1] (1,10.31) -- (60.51,10.31) -- (60.51,269.56) -- (1,269.56) -- cycle;
        \node[align=center, scale=0.6] at (30.76,139.94) {original\\queries};
        \draw [fill=white, fill opacity=1] (110.51,60.19) -- (210.51,60.19) -- (210.51,100.19) -- (110.51,100.19) -- cycle;
        \node[align=center, scale=0.6] at (160.51,80.19) {per-user contribution\\normalization count\\$<$day, region$>$};
        \draw [fill=white, fill opacity=1] (110.51,180.19) -- (210.51,180.19) -- (210.51,220.19) -- (110.51,220.19) -- cycle;
        \node[align=center, scale=0.6] at (160.51,200.19) {per-user contribution\\symptom count\\$<$day, sym., region$>$};
        \draw [fill=white, fill opacity=1] (261.51,30.19) -- (360.51,30.19) -- (360.51,70.19) -- (261.51,70.19) -- cycle;
        \node[align=center, scale=0.6] at (311.01,50.19) {daily normalization\\count\\$<$day, region$>$};
        \draw [fill=white, fill opacity=1] (261.51,90.19) -- (360.51,90.19) -- (360.51,130.19) -- (261.51,130.19) -- cycle;
        \node[align=center, scale=0.6] at (311.01,110.19) {weekly normalization\\count\\$<$week, region$>$};
        \draw [fill=white, fill opacity=1] (261.51,150.19) -- (360.51,150.19) -- (360.51,190.19) -- (261.51,190.19) -- cycle;
        \node[align=center, scale=0.6] at (311.01,170.19) {daily symptom count\\$<$day, sym., reg.$>$};
        \draw [fill=white, fill opacity=1] (261.51,210.19) -- (360.51,210.19) -- (360.51,250.19) -- (261.51,250.19) -- cycle;
        \node[align=center, scale=0.6] at (311.01,230.19) {weekly symptom count\\$<$week, sym., reg.$>$};
        \draw [fill=white, fill opacity=1] (410.51,30.19) -- (510.51,30.19) -- (510.51,70.19) -- (410.51,70.19) -- cycle;
        \node[align=center, scale=0.6] at (460.51,50.19) {noisy daily\\normalization count\\$<$day, region$>$};
        \draw [fill=white, fill opacity=1] (410.51,90.19) -- (510.51,90.19) -- (510.51,130.19) -- (410.51,130.19) -- cycle;
        \node[align=center, scale=0.6] at (460.51,110.19) {noisy weekly\\normalization count\\$<$week, region$>$};
        \draw [fill=white, fill opacity=1] (410.51,180.19) -- (510.51,180.19) -- (510.51,220.19) -- (410.51,220.19) -- cycle;
        \node[align=center, scale=0.6] at (460.51,200.19) {noisy daily/weekly\\symptom count\\$<$d./w., sym. reg.$>$};
        \draw [fill=white, fill opacity=1] (560.51,120.19) -- (659.51,120.19) -- (659.51,160.19) -- (560.51,160.19) -- cycle;
        \node[align=center, scale=0.6] at (610.01,140.19) {published\\ daily/weekly metric\\$<$d./w., sym. reg.$>$};
        %Straight Arrows
        \draw (60.51,80.31) -- (107.51,80.31);
        \draw [shift={(110.51,80.31)}, rotate=180] [fill={rgb, 255: red, 0; green, 0; blue, 0}] [line width=0.08] [draw opacity=0] (8.93,-4.29) -- (0,0) -- (8.93,4.29) -- cycle;
        \node[align=center, scale=0.6] at (85.51,56.19) {extract\\and\\bound};
        \draw (60.51,200.31) -- (107.51,200.31);
        \draw [shift={(110.51,200.31)}, rotate=180] [fill={rgb, 255: red, 0; green, 0; blue, 0}] [line width=0.08] [draw opacity=0] (8.93,-4.29) -- (0,0) -- (8.93,4.29) -- cycle;
        \node[align=center, scale=0.6] at (85.51,176.19) {extract\\and\\bound};
        \draw (360.51,49.31) -- (407.51,49.31);
        \draw [shift={(410.51,49.31)}, rotate=180] [fill={rgb, 255: red, 0; green, 0; blue, 0}] [line width=0.08] [draw opacity=0] (8.93,-4.29) -- (0,0) -- (8.93,4.29) -- cycle;
        \node[align=center, scale=0.6] at (384.51,38.19) {add noise};
        \draw (360.51,110.31) -- (407.51,110.31) ;
        \draw [shift={(410.51,110.31)}, rotate=180] [fill={rgb, 255: red, 0; green, 0; blue, 0}] [line width=0.08] [draw opacity=0] (8.93,-4.29) -- (0,0) -- (8.93,4.29) -- cycle;
        \node[align=center, scale=0.6] at (384.51,98.19) {add noise};
        %Curve Arrows
        \draw (210.51,80.31) .. controls (244.05,80.07) and (227.98,51.6) .. (258.54,50.12);
        \draw [shift={(261.51,50.06)}, rotate=180] [fill={rgb, 255: red, 0; green, 0; blue, 0}] [line width=0.08] [draw opacity=0] (8.93,-4.29) -- (0,0) -- (8.93,4.29) -- cycle;
        \node[align=center, scale=0.6] at (217.51,36.19) {aggregate\\across users};
        \draw (210.51,201.31) .. controls (244.05,201.07) and (227.98,171.3) .. (258.54,169.75);
        \draw [shift={(261.51,169.69)}, rotate=180] [fill={rgb, 255: red, 0; green, 0; blue, 0}] [line width=0.08] [draw opacity=0] (8.93,-4.29) -- (0,0) -- (8.93,4.29) -- cycle;
        \node[align=center, scale=0.6] at (217.51,155.19) {aggregate\\across users};
        \draw (210.51,80.31) .. controls (244.05,80.07) and (227.17,107.82) .. (257.68,109.71);
        \draw [shift={(260.65,109.81)}, rotate=180] [fill={rgb, 255: red, 0; green, 0; blue, 0}] [line width=0.08]  [draw opacity=0] (8.93,-4.29) -- (0,0) -- (8.93,4.29) -- cycle;
        \draw (210.51,201.31) .. controls (244.05,201.07) and (227.98,228.11) .. (258.54,229.96);
        \draw [shift={(261.51,230.06)}, rotate=180] [fill={rgb, 255: red, 0; green, 0; blue, 0}] [line width=0.08] [draw opacity=0] (8.93,-4.29) -- (0,0) -- (8.93,4.29) -- cycle;
        \draw (360.51,170.06) .. controls (393.4,169.82) and (377.64,197.68) .. (407.6,199.59);
        \draw (360.51,230.31) .. controls (393.4,230.07) and (377.64,201.25) .. (407.6,199.74);
        \draw [shift={(410.51,199.69)}, rotate=180] [fill={rgb, 255:red, 0; green, 0; blue, 0}] [line width=0.08] [draw opacity=0] (8.93,-4.29) -- (0,0) -- (8.93,4.29) -- cycle;
        \node[align=center, scale=0.6] at (402.51,153.19) {decide d./w.\\and add noise};
        \draw (510.51,49.06) .. controls (549.51,48.82) and (521.97,135.2) .. (557.64,139.75);
        \draw (510.51,110.06) .. controls (549.51,109.82) and (521.97,138.21) .. (557.64,139.86);
        \draw (510.51,200.06) .. controls (549.51,199.82) and (521.97,142.65) .. (557.64,140.03);
        \draw [shift={(560.51,139.94)}, rotate=180] [fill={rgb, 255: red, 0; green, 0; blue, 0}] [line width=0.08] [draw opacity=0] (8.93,-4.29) -- (0,0) -- (8.93,4.29) -- cycle;
        \node[align=center, scale=0.6] at (567.51,45.19) {compute metric and\\filter unreliable\\result};
    \end{tikzpicture}
    \caption{System diagram of the data generation and anonymization process}
    \label{fig:sysdiag}
\end{figure}
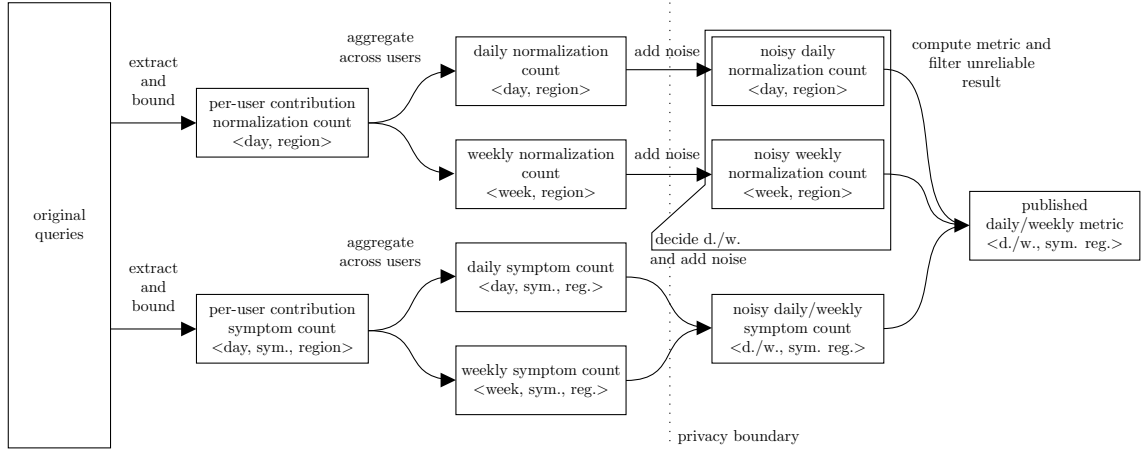

The COVID-19 Search Trends symptoms dataset is based on anonymized counts designed to protect personal data. More precisely, we protect every user’s symptom search activity on a particular day with $\varepsilon$-differential privacy for $\varepsilon = 1.68$. The majority of this $\varepsilon$, $97.5\%$, is used to anonymize symptom search counts, while $2.5\%$ is used to anonymize general search activity for normalization purposes. See Figure~\ref{fig:sysdiag} for a system diagram of the anonymization process. All counts are anonymized by adding appropriately scaled Laplace noise~\cite{laplace}. We generate this noise using our open-source differential privacy library~\cite{dplib}.

\subsection{Symptom Search Counts}
To generate the symptoms dataset, we count the number of searches that relate to a particular medical symptom and group them by date and geographical region. As a result, each count corresponds to a triplet of $<$day, symptom, region$>$.
For each day and geographic granularity level, a user can contribute at most once to any given count (per-symptom bound) and to no more than three counts in total (cross-symptom bound). Similar to the process described by Wilson et al.~\cite{privatesql}, we arbitrarily discard symptom searches of a user that exceed their contribution bound. Approximately $75\%$ of users search at most three symptoms per day, so the amount of data dropped is relatively small.

As an example, assume that on June 3, 2020 a user made two searches related to fever in Santa Clara county (CA), one search related to fever in San Bernardino county (CA), one search related to fever in Clark county (NV), and one search related to cough in Clark county (NV). The following table lists a possible configuration of the user’s contribution to the respective counts after contribution bounding.

\begin{table}[H]
    \centering
    \begin{tabular}{| l | l | l | l |}
        \hline
        level   & count                                     & contribution      & bound type \\
        \hline
        0       & $<$2020-06-03, fever, United States$>$    & 1 (originally 4)  & per-symptom \\
        0       & $<$2020-06-03, cough, United States$>$    & 1                 & \\
        1       & $<$2020-06-03, fever, California$>$       & 1 (originally 3)  & per-symptom \\
        1       & $<$2020-06-03, fever, Nevada$>$           & 1                 & \\
        1       & $<$2020-06-03, cough, Nevada$>$           & 1                 & \\
        2       & $<$2020-06-03, fever, Santa Clara$>$      & 1 (originally 2)  & per-symptom \\
        2       & $<$2020-06-03, fever, San Bernardino$>$   & 1                 & \\
        2       & $<$2020-06-03, fever, San Bernardino$>$   & 1                 & \\
        2       & $<$2020-06-03, fever, Clark$>$            & 1                 & \\
        2       & $<$2020-06-03, cough, Clark$>$            & 0 (originally 1)  & cross-symptom \\
        \hline
  \end{tabular}
  \label{tab:contribbound}
  \caption{Example of per-symptom and cross-symptom contribution bounding}
\end{table}

To protect a user’s daily symptom searches with $1.638$-differential privacy ($97.5\%$ of the total $\varepsilon$), we then add appropriately scaled Laplace noise. In the case of daily granularity, the noise is added directly to the respective count. In the case of weekly granularity, we first sum the daily counts to obtain a weekly count and then add the noise to the weekly count. The following table lists the noise parameterization for each granularity level in terms of its scale b and standard deviation $\sigma = \sqrt{2}b$. The total $\varepsilon$ sums up to $1.638$, i.e., the privacy budget we spend on computing symptom searches.

\begin{table}[H]
    \centering
    \begin{tabular}{| l | l | l |}
        \hline
        level   & noise added to daily or weekly count                                  &  $\varepsilon$ per level \\
        \hline
        0       & $b = 3 / \varepsilon_0 \approx 17.857 \quad (\sigma \approx 25.254)$  &  $\varepsilon_0 = 0.168$  \\
        1       & $b = 3 / \varepsilon_1 \approx 8.108 \quad (\sigma \approx 11.467)$   &  $\varepsilon_1 = 0.37$  \\
        2       & $b = 3 / \varepsilon_2 \approx 2.727 \quad (\sigma \approx 3.857)$    &  $\varepsilon_2 = 1.1$  \\
        \hline
  \end{tabular}
  \label{tab:symnoise}
  \caption{Noise parameters for daily and weekly symptom search counts}
\end{table}

\subsection{Normalization Counts}
To normalize our data, we scale each daily or weekly symptom search count proportional to the total search activity in the given geographical region during the respective time period. We estimate the search activity based on the number of unique users who have issued a search query in a particular geographical region during a day. Consequently, each daily normalization count corresponds to a tuple of $<$day, region$>$.

For each day and each geographic granularity level, a user can contribute to at most one normalization count. Similarly to the symptom search counts, we arbitrarily discard contributions that exceed this contribution limit.

The daily and weekly normalization counts are protected with 0.021-differential privacy each, or $0.042$-differential privacy in combination ($2.5\%$ of the total $\varepsilon$). In the case of daily counts, we add appropriately scaled Laplace noise to the respective count. In the case of weekly counts, we first sum the respective daily counts and then add Laplace noise to the sum. The following table lists the noise parameterization for each granularity level in terms of its scale b and standard deviation $\sigma = \sqrt{2}b$. Note that the total privacy budget sums up to $0.042$.

\begin{table}[H]
    \centering
    \begin{tabular}{| l | l | l |}
        \hline
        level   & noise added to daily and weekly count                                     & $\varepsilon$ per level \\
        \hline
        0       & $b = 1 / \varepsilon'_0 \approx 434.783 \quad (\sigma \approx 614.875)$   &  $\varepsilon'_0 = 0.0023$  \\
        1       & $b = 1 / \varepsilon'_1 \approx 212.766 \quad (\sigma \approx 300.897)$   &  $\varepsilon'_1 = 0.0047$  \\
        2       & $b = 1 / \varepsilon'_2 \approx 71.429 \quad (\sigma \approx 101.015)$    &  $\varepsilon'_2 = 0.014$  \\
        \hline
  \end{tabular}
  \label{tab:normnoise}
  \caption{Noise parameters for daily and weekly normalization counts}
\end{table}

\section{Reporting the Metrics}
The daily or weekly data published in the COVID-19 Search Trends symptoms dataset are based exclusively on the anonymized counts described in the previous section. Due to the post-processing property of differential privacy, the following process does not consume any privacy budget.

\paragraph{Computing the Reported Data}
Given a particular geographic region, symptom and time interval (either day or week), the normalized search count published in the COVID-19 Search Trends is computed as 
\[c \cdot \max\{(A / B), 0\}\], 
where $A$ is the noisy symptom search count, $B$ is the normalization count and $c$ is a scaling factor specific to the geographic region. 
For each geographic region, $c$ is chosen in a way that maps the data in the initial release of the symptoms dataset to values between 0 and 100. Because we keep $c$ fixed, future releases may contain metrics that exceed the value of 100. We also want to note that $c$ is determined purely based on noisy symptom search counts and normalization counts, so no privacy budget is spent on its computation.

\paragraph{Removing Unreliable Data}
In some geographic regions, the noise added for differential privacy reasons can introduce a disproportionate amount of uncertainty to a metric. Typically, this happens when the respective symptom search count is empty or small. To address this uncertainty, we only keep a metric if it has a chance of $50\%$ or more to be within $25\%$ points of its raw value, i.e., the metric before adding noise.

More precisely, let $A$ be some noisy symptom search count obtained after adding Laplace noise to the raw symptom search count $a^*$ (note that $a^*$ is not noisy but still subject to contribution bounding). Similarly, let $B$ be the corresponding noisy normalization count computed from the raw count $b^*$. To decide whether the data associated with $A$ will be kept or dropped we:

\begin{itemize}
    \item Compute a confidence interval $[l, r]$ based on $A$ and $B$ that contains $a^* / b^*$ with a probability of at least $50\%$.
    \item Keep the data if $|A / B - l| \leq 0.25 \cdot A / B$ and $|A / B - r| \leq 0.25 \cdot A / B$. Otherwise drop it.
\end{itemize}

The confidence interval $[l, r]$ is entirely based on the anonymized counts $A$ and $B$. As a result, no privacy budget is spent on removing unreliable data.

\paragraph{Deciding Between Daily and Weekly Granularity}
The decision whether a certain symptom is published at the daily or weekly granularity within a given geographic region is based on the amount of data for this symptom in that region. Intuitively, if metrics are available for more than half of the days within the sample period from February 2020 to July 2020 (i.e., not dropped as unreliable, as explained in the previous section), we want to produce data for this symptom at the daily resolution. Otherwise, if too many daily metrics are dropped, we opt for the weekly resolution, which is more reliable.

For the sake of usability, we make the decision between daily and weekly granularity once and stick to it over the course of the entire data release.

To avoid consuming additional privacy budget, we switch between daily and weekly granularity based on the anonymized volume of search activity. More precisely, for each symptom we make our decision according to the following process:

\begin{itemize}
    \item All level 0 regions are published with daily granularity.
    \item We order all other regions that are of the same level and contained in the same level 0 region by the total search activity, which we approximate based on the anonymized normalization counts.
    \item Starting with the geographic region that has the highest search activity, we start publishing daily metrics according to the process described above.
    \item We continue down the list of regions, by decreasing volume of search activity. For each new region, we look at the last 20 regions we published\footnote{If less than 20 regions were published, we only consider the regions published so far.}. If 11 or more had more than $50\%$ of metrics dropped in the time period from February 2020 to July 2020, we switch from daily to weekly granularity and publish weekly metrics for all regions from this point forward. Otherwise, we publish the current region with daily granularity, and repeat the process.
\end{itemize}

Taking a majority vote over the last 20 regions mitigates potential outliers in the ordering.

The key insight is that this process uses anonymised data only to order the regions, i.e., the daily normalization counts. Thus the privacy budget we spend on the ordering is already accounted for. Moreover, we never compute anonymized daily and weekly symptom search counts for the same symptom and region pair at the same time. This means we only spend privacy budget on one of the two counts, resulting in the promised $\varepsilon = 1.68$. 

\bibliographystyle{unsrt}
\bibliography{references}
\end{document}